\documentclass{article}\usepackage{jkas2}
\runningauthor{Feretti et al.}
\runningtitle{Properties and Spectral Behaviour of Cluster Radio Halos}
\beginpage{1}
\endpage{6}

\begin{document}

\font\twelvei = cmmi10 scaled\magstep1 
       \font\teni = cmmi10 \font\seveni = cmmi7
\font\mbf = cmmib10 scaled\magstep1
       \font\mbfs = cmmib10 \font\mbfss = cmmib10 scaled 833
\font\msybf = cmbsy10 scaled\magstep1
       \font\msybfs = cmbsy10 \font\msybfss = cmbsy10 scaled 833
\textfont1 = \twelvei
       \scriptfont1 = \twelvei \scriptscriptfont1 = \teni
       \def\mit{\fam1 }
\textfont9 = \mbf
       \scriptfont9 = \mbfs \scriptscriptfont9 = \mbfss
       \def\bmit{\fam9 }
\textfont10 = \msybf
       \scriptfont10 = \msybfs \scriptscriptfont10 = \msybfss
       \def\bmsy{\fam10 }

\def\etal{{\it et al.~}}
\def\eg{{\it e.g.,~}}
\def\ie{{\it i.e.,~}}
\def\lsim{\raise0.3ex\hbox{$<$}\kern-0.75em{\lower0.65ex\hbox{$\sim$}}}
\def\gsim{\raise0.3ex\hbox{$>$}\kern-0.75em{\lower0.65ex\hbox{$\sim$}}}
\def\ltsim{\hbox{\raise 2pt \hbox {$<$} \kern-1.1em \lower 4pt \hbox {$\sim$}}}
\def\ltapprox{\hbox{\raise 2pt \hbox {$<$} \kern-1.1em \lower 5pt \hbox 
{$\approx$}}}
\def\gtsim{\hbox{\raise 2pt \hbox {$>$} \kern-1.1em \lower 4pt \hbox {$\sim$}}}
\def\gtapprox{\hbox{\raise 2pt \hbox {$>$} \kern-1.1em \lower 5pt \hbox 
{$\approx$}}}
\def\arcsec{$^{\prime\prime}$}
\def\arcmin{$^{\prime}$}
\def\degrees{$^{\circ}$}

\title{Properties and Spectral Behaviour of Cluster Radio Halos}

\author{L. Feretti$^{1}$,  G. Brunetti$^{1}$,
G. Giovannini$^{1,2}$,  N. Kassim$^{3}$, E. Orr\'u$^{4}$, G.Setti$^{1,2}$}
\address{$^{1}$ Istituto di Radioastronomia,
Via P. Gobetti 101, Bologna, Italy}
%{\it E-mail: lferetti@ira.cnr.it}
\address{$^{2}$ Dipartimento di Astronomia, Univ. Bologna, via Ranzani 1, 
Bologna, Italy}
\address{$^{3}$ Naval Research Laboratory, Code 7213, Washington, DC, 
20375 USA }
\address{$^{4}$ Dipartimento di Fisica, Univ. Cagliari, 
and INAF-Oss. Astron. di Cagliari, Cagliari, Italy
}

%\address{\normalsize{\it (Received ... Accepted ...)}}

\abstract{ Several arguments have been presented in the literature to
support the connection between radio halos and cluster mergers.  The
spectral index distributions of the halos in A665 and A2163 provide a
new strong confirmation of this connection, i.e. of the fact that the
cluster merger plays an important role in the energy supply to the
radio halos.  Features of the spectral index (flattening and patches)
are indication of a complex shape of the radiating electron spectrum,
and are therefore in support of electron reacceleration models.
Regions of flatter spectrum are found to be related to the recent
merger.  In the undisturbed cluster regions, instead, the spectrum
steepens with the distance from the cluster center.  
The plot of the integrated spectral index
of a sample of halos versus the  cluster temperature indicates
that clusters at higher temperature tend to host halos with flatter
spectra. This correlation provides further evidence
of the connection between radio emission and
cluster mergers.}

\keywords{Radio emission - X-ray emission - Clusters of Galaxies -
Intergalactic medium  }

\maketitle

\section {Introduction}

Radio halos are the most spectacular expression of cluster non--thermal
emission. They are low brightness extended radio sources permeating
the cluster centers, similarly to the X-ray emitting gas, with sizes of
more than a Mpc.  The prototype of this class is Coma C (Fig. 1), the
halo source in the Coma cluster, which was first shown to be diffuse
by Willson (1970) and mapped later at various radio wavelengths by
several authors (Giovannini et al. 1993, Thierbach et al. 2003, and
references therein).

Studies of several radio halos and of their hosting clusters have been
recently performed, thus improving the knowledge of the
characteristics and physical properties of this class of radio
sources.  Giant radio halos have been detected in A665 (Giovannini \&
Feretti 2000), A2163 (Feretti et al. 2001), A2219 (Bacchi et
al. 2003), A2255 (Feretti et al. 1997a), A2319 (Feretti et al. 1997b),
A2744 (Govoni et al. 2001a), 1E$0657-56$ (Liang et
al. 2000), and CL$0016+16$ (Giovannini \& Feretti 2000).  The latter
cluster, at z = 0.555, is the most distant cluster with a radio halo
known so far. Radio halos of small size, i.e. $<<$ 1 Mpc, have also
been revealed in the central regions of some clusters, as
in A401 (Giovannini \& Feretti 2000), A1300 (Reid et al. 1999),
A2218 (Giovannini \& Feretti 2000) and A3562 (Venturi et al. 2003).

In this paper we outline the general properties of radio halos, and we
focus on the spectral index maps recently obtained for the
halos in the clusters A665 and A2163. 
Spectral index maps represent a powerful tool to study the
properties of the relativistic electrons and of the magnetic field in
which they emit, and to investigate the connection between the
electron energy distribution and the intracluster medium (ICM).

\begin {figure}[t]
\vskip 0cm
\centerline{\epsfysize=9.5cm\epsfbox{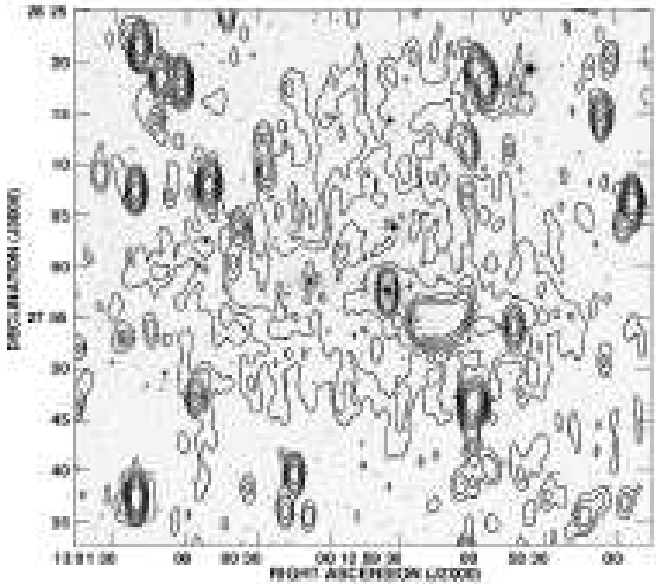}}
\vskip -0.2cm
\label{fig1}
\caption{
 Isocontour map at 0.3 GHz of the central region of the Coma cluster
superimposed onto the optical image from the DPSS.  
The resolution of the radio image is 55\arcsec$\times$ 125\arcsec 
(FWHM, RA $\times$ DEC); contour levels are: 3, 6, 12, 25, 50, 100 mJy/beam.}
\vskip -0.5cm
\end{figure}

The intrinsic parameters quoted in this paper are
computed with a Hubble constant H$_0$ = 70 km s$^{-1}$ Mpc$^{-1}$ and
a deceleration parameter q$_0$ = 0.5.

\section {General properties}

The general properties of radio halos, derived from observational
data, can be summarized as follows:

a) Under the standard equipartition conditions, assuming equal energy in
relativistic protons and electrons (k = 1) and a volume filling factor
$\Phi$ = 1, the minimum energy density in halos is of the order of
$10^{-14}-10^{-13}$ erg cm$^{-3}$, i.e. much lower than the energy
density in the thermal gas.  The corresponding equipartition magnetic
field strengths range from $\simeq$ 0.1 to 1 $\mu$G (e.g.
Govoni \& Feretti 2004).

b) The spectra of halos are steep, as typically found in aged radio
sources ($\alpha$ \gtsim~ 1, assuming $S_{\nu}\propto \nu ^{-\alpha}$).  
Steepening at high frequencies is reported in
most cases with adequate spectral coverage; e.g. in
Coma (Thierbach et al. 2003), A754 (Bacchi et al. 2003), A1914
(Komissarov \& Gubanov 1994), A2319 (Feretti et al. 1997b).  The 
radiative lifetime of the relativistic electrons, derived from
the integrated spectra considering synchrotron and inverse Compton
energy losses, is of the order of $\sim$ 10$^8$ yr.  This implies
that, to allow for the large sizes of the radio emitting regions, the
radiating particles need to be reaccelerated by some mechanism,
acting with an efficiency comparable to the energy loss processes.

c) Radio properties of halos are linked to  properties of the host
clusters.
Indeed, the radio power of a halo correlates with the cluster
X-ray luminosity (Bacchi et al. 2003), the thermal gas
temperature (Liang et al. 2000), and the total cluster mass (Govoni et 
al. 2001a). 
Moreover, in a number of well--resolved clusters, a
point--to--point spatial correlation is observed between the radio
brightness of the halo and the X-ray brightness as detected by {\it
ROSAT} (Govoni et al. 2001b).  This correlation is visible e.g. in
A2744 also in the {\it Chandra} high resolution data (Kempner \& David
2004).

d) The detection rate of radio halos in a complete X-ray
flux--limited sample of clusters is $\simeq 5$\% (at the detection limit of the
NRAO VLA Sky Survey).  The halo fraction increases with the X-ray
luminosity, up to $\simeq 33$\% for clusters with $L_X> 10^{45}$ erg
s$^{-1}$.  Thus, halos are present in rich clusters, characterized by high
X-ray luminosities and temperatures (Giovannini \& Feretti 2002).  

e) Halos are typically found in clusters showing distorted X-ray morphology 
and significant substructure (Schuecker et al. 2001, Buote 2001), and strong
gas temperature gradients (Govoni et al. 2004). Some clusters show a
spatial correlation between the radio halo brightness and the hot gas
regions, although this is not a general feature.  Thus, radio halos
are strictly related to the presence of cluster merger processes.
According to recent suggestions, a recent major cluster merger event
is the relevant factor for the formation of a radio halo
by supplying the energy
for the reacceleration of radiating electrons (e.g., Feretti 2003
and references therein).

\begin {figure}[t]
\vskip 0cm
\centerline{\epsfysize=9.cm\epsfbox{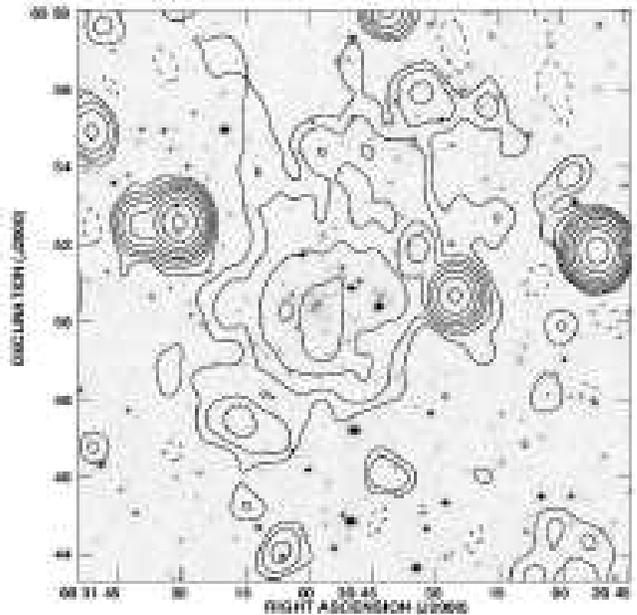}}
\vskip -0.2cm
\label{fig2}
\caption{
 Isocontour map at 1.4 GHz of the central region of A665
superimposed onto the optical image from the DPSS.  
The resolution is 42\arcsec $\times$ 52\arcsec 
(FWHM, RA $\times$ DEC). Contour levels are: -0.2, 0.2, 0.4, 0.8, 
1.5, 3, 6, 12, 25 mJy/beam. }
\vskip -0.5cm
\end{figure}

\section{Spectral index maps}

The distribution of the radio spectral index is an important
observable in a radio halo, since it is related to the shape of
the electron energy distribution and to the properties of the magnetic
field in which they emit.  By combining high resolution spectral
information and X-ray images it is possible to investigate the
connection between the thermal and relativistic plasma both
on small scales (e.g., spectral index variations vs. clumps in the ICM
distribution) and on large scales (e.g. radial spectral index trends).

\begin {figure*}[t]
\vskip -0.6cm
\centerline{\epsfysize=10.0cm\epsfbox{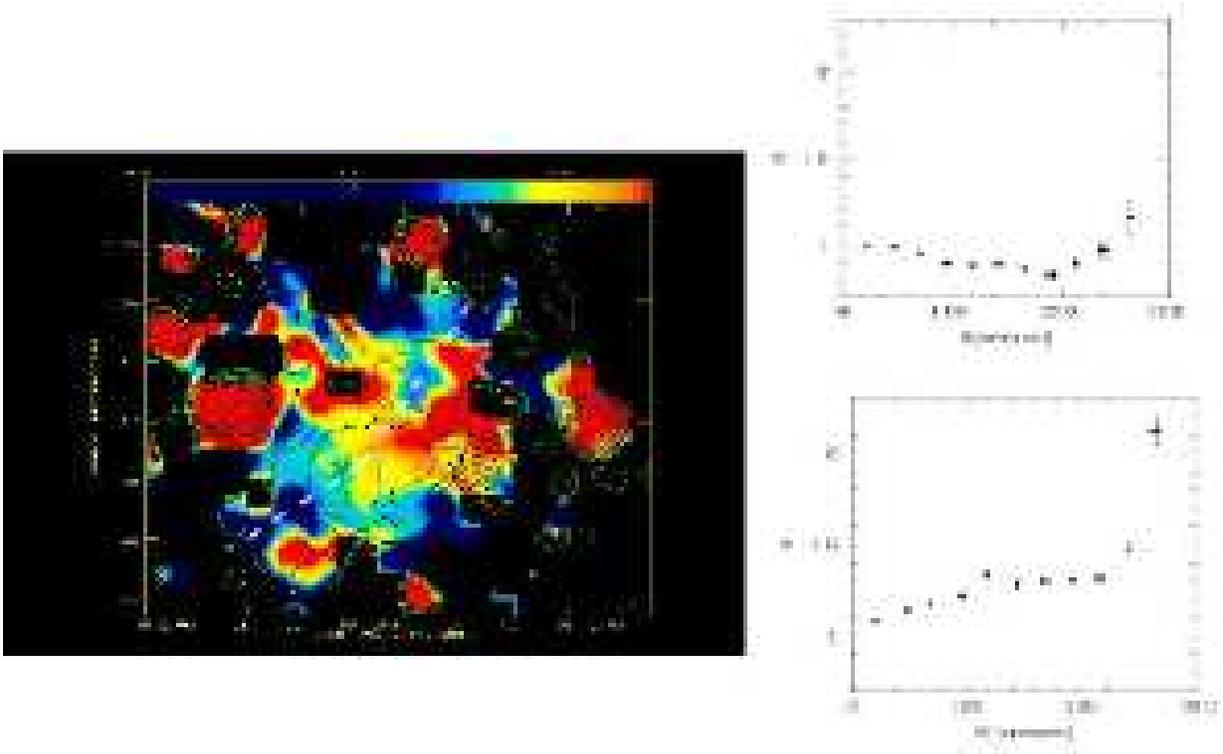}}
\vskip 0.5cm
\label{fig4}
\caption{
{\bf Left panel}: color--scale image of the spectral index
between 0.3 GHz and 1.4 GHz of A665, obtained with a resolution of
68\arcsec $\times$ 59\arcsec (PA= 25\degrees) FWHM.  The contours
indicate the radio emission at 1.4 GHz (Fig. 2).
{\bf Right panels}:
radial profiles of the spectral index along the two directions
indicated by the dashed lines in the spectral index map.
The cluster core radius, derived from X-ray data, is 96\arcsec, corresponding
to 270 kpc.
}
\vskip -0.5cm
\end{figure*}

The first spectral index image of a radio halo has been obtained by
Giovannini et al. (1993) for Coma C, using data at 0.3 GHz from the
Westerbork Synthesis Radio Telescope (WSRT) and data at 1.4 GHz from
the Very Large Array (VLA)
 and the Dominion Radio Astronomy Observatory. The image shows
a flatter spectrum in the central region ($\alpha\simeq$ 0.8) and a progressive
steepening with increasing distance from the center (up to
$\alpha\simeq$ 1.8 at a distance of about 15\arcmin).  This trend
is confirmed by a new spectral index map derived by comparing the 1.4
GHz image obtained with the Effelsberg single dish by Deiss et al. (1997),
and the 0.3 GHz image obtained from the combination of VLA and WSRT
data (Giovannini et al. in preparation).  The high sensitivity of the images
allows the computation of the spectral index up to $\sim$ 30\arcmin~
from the cluster center, where the spectral index is $\alpha\simeq$ 2.
The high resolution X-ray data of the Coma cluster  obtained with
XMM-Newton provide the evidence of recent merger activity at scales larger
than 10\arcmin, whereas the cluster core is suggested to be in a
basically relaxed state (Arnaud et al. 2001).

New spectral index images have been recently obtained for the clusters
A665 and A2163 using VLA data at 0.3 and 1.4 GHz. 
They are presented below. For a more detailed
analysis see Feretti et al. (2004). 

{\bf A~665} (z = 0.1818).  The radio halo in this cluster is shown in
Fig. 2 (Giovannini \& Feretti 2000).  The radio emission, of $\sim$
1.3 h$_{70}^{-1}$ Mpc in size, is asymmetric with respect to
the cluster center, being brighter and more extended towards the  NW. 
This is the
direction where the X-ray brightness distribution detected from
Chandra data is elongated, probably due to the merging of a smaller
subcluster (Markevitch \& Vikhlinin 2001).  The Chandra data also
reveal the presence of a remarkable shock in front of the cool cluster
core, indicating that the core is moving with a relatively high Mach
number.  The shock is located near the southern boundary of the radio
halo.  The complex temperature structure is confirmed by more
sensitive data published by Govoni et al. (2004).

The spectral index map is clumpy (Fig. 3). The spectrum
in the central halo region is rather constant, with spectral index
values between 0.8 and 1.2 within one core radius from the cluster
center (i.e. within $\sim$ 96\arcsec).  In the northern region of
lower radio brightness the spectrum is flatter than in the southern
halo region.

Starting from the approximate radio peak position, we obtained
profiles of the spectral index trend by averaging the values of the
spectral index within small sectors around the two directions marked
by dashed lines in Fig. 3.  The spectrum in the NW direction flattens up to a
distance of about 200\arcsec~ from the center (Fig. 3, top right 
panel).  This is the region where the asymmetric extended
X-ray emission indicates the existence of an ongoing
merger with another cluster.  The gas temperature in this region
(Markevitch \& Vikhlinin 2001, Govoni et al. 2004) shows strong
variations, from about 12 keV in the NE to about 8 keV in the
SW. The spectral index flattening follows the X-ray morphology, but
there is no one--to--one correspondence with the value of the gas
temperature.  Therefore, it seems that this region, which is 
strongly influenced by the merger, is a shocked region, where the gas
at different temperatures is still in the process of mixing.

In defining a profile in the southern cluster region, we tried to
avoid the region where there is a possible contamination of discrete
sources.  The spectrum in this direction (Fig. 3, bottom right panel) 
steepens significantly from the center to the periphery.
The spectral index  increase from $\alpha$ $\sim$ 1 to $\alpha$ \gtsim~
2 is gradual at the beginning, and occurs on a scale of less than 3
cluster core radii.  The region of constant spectral index at distance
between 120\arcsec~ and 220\arcsec~ from the center is located NE of
a discrete source and could be affected by  its presence.
We note that this path crosses the region of the hot shock
detected by Chandra at the southernmost edge of the radio halo
(Markevitch \& Vikhlinin 2001). The shock is located at about
100\arcsec~ from the center along the profile. No significant spectral
flattening is detected at this position.

{\bf A~2163} (z = 0.203).  This cluster is one of the hottest and most X-ray
luminous among known clusters. It  hosts a powerful radio halo,
extended $\sim$ 2 h$_{70}^{-1}$ Mpc (Fig. 4), 
studied in detail by Feretti et al. (2001).

\begin {figure}[t]
\vskip 0cm
\centerline{\epsfysize=8.5cm\epsfbox{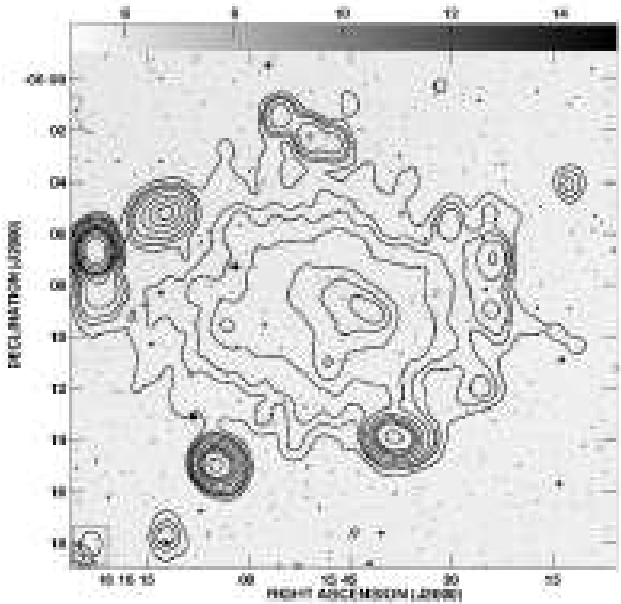}}
\vskip -0.2cm
\label{fig3}
\caption{ Isocontour radio map at 1.4 GHz of A2163
superimposed onto the optical image from the DPSS.
The resolution is 60\arcsec $\times$ 45\arcsec (FWHM, RA $\times$ DEC).
Contours are 0.2, 0.5, 0.8, 1.5, 3.0, 5.0, 7.0, 9.0, 15.0 mJy/beam.
}
\vskip -0.5cm
\end{figure}

X-ray ROSAT and ASCA data (Elbaz et al. 1995, Markevitch et al.  1996)
suggest that this cluster is likely to be have experienced 
a recent merger.  The X-ray
brightness distribution indicates elongation in the E-W direction,
suggesting that this could be the merger direction.  Recent Chandra
data (Markevitch \& Vikhlinin 2001, Govoni et al. 2004) reveal a
complex morphology indicating that the cluster central region is in a
state of violent merger.  The temperature map is also complex, with
variations by at least a factor of 2, suggesting streams of hot and
cold gas flowing in different directions, as well as a possible
remnant of a cool gas core, surrounded by shock--heated gas.  The
structure in the temperature map is too complicated to easily infer
the geometry of the merger.

The spectral index map (Fig. 5, left panel) is clumpy in the central
cluster region, but it is rather constant within one core radius,
showing spectral index values between 1 and 1.1.  On a larger scale,
there is evidence that the western halo region is flatter than the
eastern one.  In particular, there is a vertical region crossing the
cluster center and showing flatter spectrum, with a clear evidence of
spectral flattening both at the northern and at the southern halo
boundaries.

Radial profiles of the spectral index along two interesting directions
(see dashed lines) have been obtained, as in A665, by averaging the
values of the spectral index within small sectors around the two
directions.  The spectral index profile along the N-S direction
(Fig. 5, top right panel) is globally rather flat and shows a
significant flattening at about 300\arcsec~ from the center (note that
the strong flattening of the two last points is due to the presence of
a discrete radio source).  In the eastern cluster region, in the area
free of discrete sources, the spectrum becomes progressively steeper
from the center to the periphery. This is well seen in the profile
along the S-E direction (bottom right panel).
The spectral steepening is weaker than in A665, and occurs
over a much larger scale.  The vertical region of flatter spectrum
coincides with the region of highest temperature detected from
Chandra, thus it  is likely related to the
strong dynamical activity at the cluster center.  The N-S extent of
the region with flat spectrum is in support of a merger occurring in
the E-W direction, as indicated by the X-ray brightness
distribution. The complexity of the merger is, however, reflected in
the complexity of the spectral index map.

\begin {figure*}[t]
\vskip -0.6cm
\centerline{\epsfysize=11.8cm\epsfbox{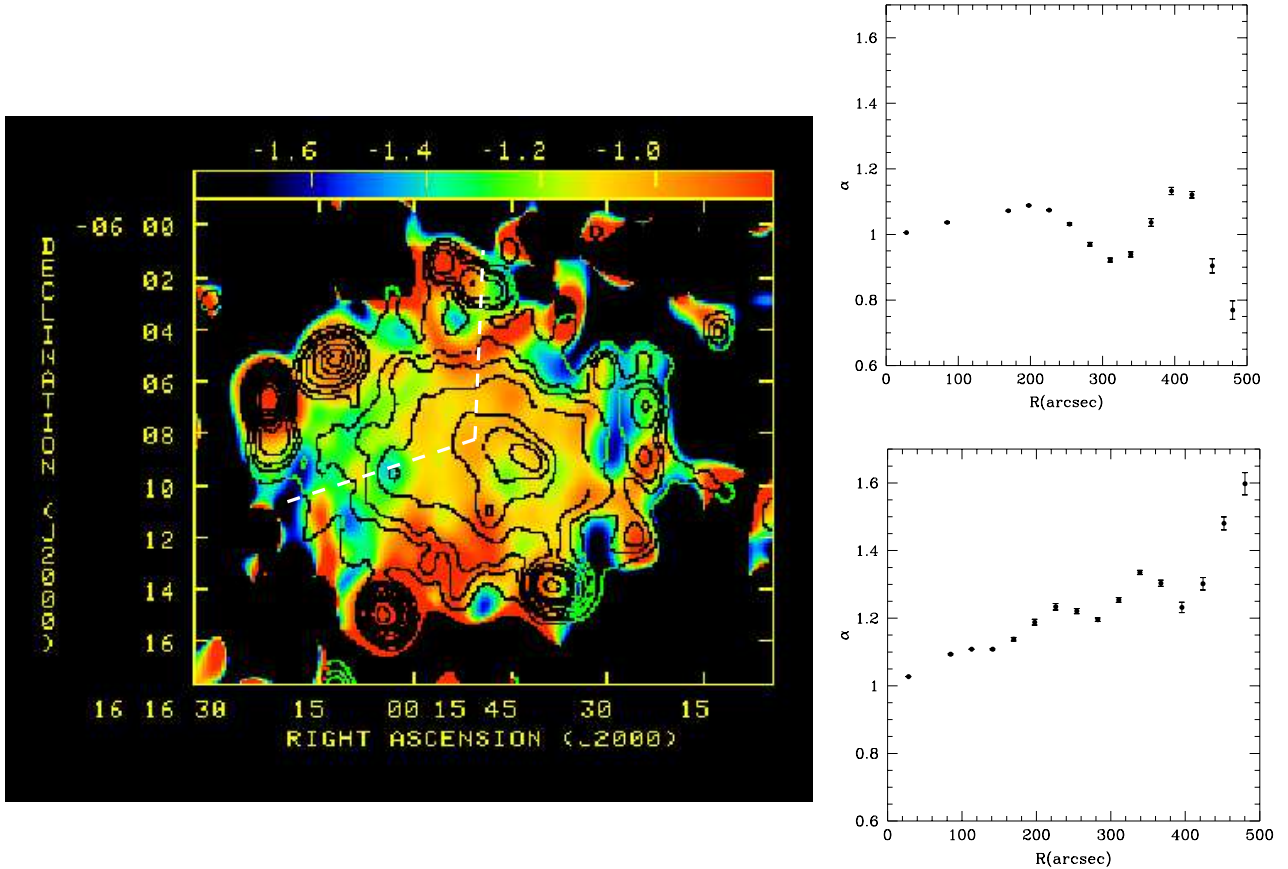}}
\vskip -0.1cm
\label{fig5}
\caption{{\bf Left panel}: color-scale image of the spectral index
between 0.3 GHz and 1.4 GHz of A2163, obtained with a resolution of
60\arcsec $\times$ 51\arcsec (PA=0\degrees) FWHM.  The contours
indicate the radio emission at 1.4 GHz (Fig. 4). 
 {\bf Right panels}:
radial profiles of the spectral index along the two directions
indicated by the dashed lines in the spectral index map.
The cluster core radius is 72\arcsec, corresponding to 220 kpc.
}
\vskip -0.5cm
\end{figure*}

\section{Connection to cluster merger}

The spectral index maps of A665 and A2163 indicate the existence of patches
of different spectra, with significant variations on
scales of the order of the observing beam ($\sim$ 200 kpc).  This
suggests a complex shape of the electron spectrum, as generally
expected in the case of particle reacceleration.

The regions influenced by an ongoing merger show a different behaviour
with respect to the relatively undisturbed regions.  Regions of
flatter spectra are indication of the presence of more energetic
radiating particles, and/or of a larger value of the local magnetic
field strength. Flatter spectra are found in regions influenced by
merger processes, while a general radial spectral steepening is found
in more undisturbed cluster regions. This behaviour is qualitatively
expected by electron reacceleration models.

We have attempted the evaluation of the energy supplied to the halos
in the regions of flatter spectral index, by matching the
reacceleration gains and the radiative losses of the radio emitting
electrons.  In regions of identical volume and  brightness at 0.3
GHz, a flattening of the spectral index from 1.3 to 0.8 implies that
the energy injected into the electron population is larger by a factor
of $\sim$2.5. 
 Since the lifetimes of 
the radio emitting electrons at $\sim$ 1 GHz in a 0.5 $\mu$G field are
of the order of $\sim$ 10$^8$ yr, the energy injection event 
should be recent. Electrons in flatter spectrum regions have a
spectral cutoff at higher energies, thus they have
been reaccelerated more recently.

These results prove that the radio spectral index can be a powerful
tracer of the current physical properties of clusters, and confirms
the importance of cluster mergers in the energetics of relativistic
particles responsible for halo radio emission.  On the other hand,
the spectral index steepens progressively with the distance from the
cluster center in the more relaxed regions.
 This is another indication that the energy of relativistic
particles is sensitive to the effect of mergers.

It is worth noticing that there is no evidence of spectral flattening
at the location of the hot shock detected in A665 (Markevitch \&
Vikhlinin 2001).  This is consistent with the fact that shocks in
major mergers are too weak to produce a significant number
of high energy particles (Gabici \& Blasi
2003, Berrington \& Dermer 2003) 
and indeed the Mach number of the shock in A665 is $\sim$ 2
(Markevitch \& Vikhlinin 2001). Our result supports the scenario that
cluster turbulence might be the major mechanism responsible for the supply of
energy to the radiating electrons.

To check if there is a more general connection between the radio halo
spectra and the cluster properties, we have investigated the existence
of a possible correlation between the spectral index of a sample of
radio halos with good spectral information and the cluster
temperature. In Fig. 6, we plot the integrated average spectral index
obtained over the available frequency range versus the average cluster
temperature derived from the X-ray data. The plot may suggest that
clusters at higher temperature tend to host halos with flatter
spectra, although the data are still scarce and the errors very large.
This correlation, if confirmed, could be understood in the framework
of the reacceleration models, since the hottest clusters, being more
massive, may statistically host more violent mergers giving rise to a
higher fraction of the turbulent energy per unit gas mass.  Thus this
correlation would provide further evidence of the connection between
radio emission and cluster mergers.

\begin {figure}[t]
\vskip 0cm
\centerline{\epsfysize=8.cm\epsfbox{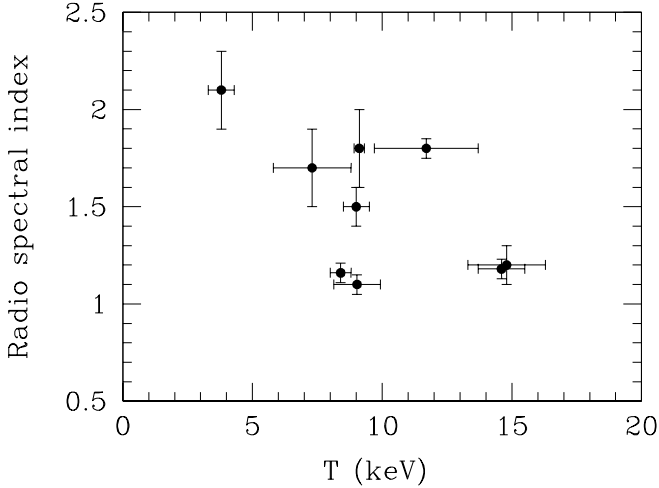}}
\vskip -0.2cm
\label{fig6}
\caption{
Integrated spectral index of radio halos versus cluster 
temperature.
}
\vskip -0.5cm
\end{figure}

\section{Implications on particle acceleration and
cluster magnetic field profile}

Since the diffusion velocity of relativistic particles is low in
relation to their radiative lifetime, the radial spectral steepening
detected in A665 and A2163 in the more undisturbed regions 
cannot be simply due to ageing of radioemitting electrons. Therefore
the spectral steepening must be related to the intrinsic evolution of
the local electron spectrum and to the radial profile of the cluster
magnetic field.

In the primary reacceleration scenario the electrons are accelerated
up to a maximum energy which is given by the balance between
acceleration efficiency and energy losses. As a consequence a break
(or cut--off) is expected in the synchrotron spectrum emitted by these
electrons.  The presence of spectral steepenings and flattenings
implies that the frequency of such a synchrotron break is relatively
close to the range 0.3--1.4 GHz for a relevant fraction of the cluster
volume.

If Fermi--like acceleration processes are efficient in the cluster
volume, the synchrotron break frequency $\nu_{\rm b}$ is related to the
magnetic field B as:

\begin{equation}
\nu_{\rm b}(r) \propto \chi^2(r) 
\cases{
B(r), & if $B(r) << B_{IC}$; \cr
                             \cr
B^{-3}(r), & if $B(r) >> B_{IC}$. \cr}
\end{equation}

\noindent
where $\chi(r)= \tau_{acc}^{-1}$ is the acceleration 
rate ($dp/dt = \chi p$), and $B_{IC}$ is the IC equivalent 
magnetic field $\sim 3 \mu$G.

The equipartition magnetic field in the present clusters are 
$\sim$ 0.5 $\mu$G, i.e.
$<< B_{IC}$, thus we expect $\nu_{\rm b}(r)
\propto \chi^2(r)B(r)$.  
Allowing for a decrease of $B$ with $r$, it follows that a roughly
constant acceleration efficiency results in a systematic steepening of
the synchrotron spectrum with $r$, simply because at a given frequency
higher energy electrons emit in the lower field intensity.  This
steepening effect would be further enhanced if the reacceleration
efficiency increases toward the central regions.

The trends of the spectral index showing radial steepening (bottom
right panels of Figs. 3 and 5) reflect the radial trends of the
product between the magnetic field strength and the reacceleration
efficiency, $B \chi^2$. In both clusters, it is derived that
$B \chi^2$ decreases about a factor  
of 2  over the scale of the spectral
index profile.  Each trend of $B \chi^2$ represents the profile of the
magnetic field strength in the case that the reacceleration is
constant throughout the cluster.  

Under the hypothesis that the
magnetic field results from the compression of the thermal plasma
during the gravitational collapse, we would expect  $B \propto
n_{th}^{2/3}$, where $n_{th}$ is the thermal plasma density.  We
derive that this trend is much steeper that 
than that estimated by the
spectral behaviour.  We note, however, that in A665 and A2163 the
$\beta$ model approximation for the distribution of $n_{th}$ may be 
significantly inaccurate owing to gas 
perturbations related to strong dynamical
evolution.  On the other hand, detailed MHD numerical simulations show
that the radial behaviour of the magnetic field may
diverge from the prediction of a frozen--in B model, resulting in flatter
spectra in the central regions, and steeper in the external regions (Dolag et
al. 1999, 2002).

The radial profile of the Coma cluster magnetic field, derived in the
same way by modeling the spectral steepening (Brunetti et al. 2001),
shows a decline with the distance which is not too different from that
expected by the frozen--in B model.  This is also the case of the
magnetic profile in A119, obtained using the Rotation Measure of the
cluster radio galaxies (Dolag et al. 2001).  In general, we could
argue that the ongoing violent mergers in A665 and A2163 are likely to
play a crucial role in determining the conditions of the radiating
particles and the magnetic field in these clusters.

We finally remark that the hypothesis of constant reacceleration
efficiency in the cluster volume may not be valid, according to the
following arguments: i) first results from numerical simulations
(Norman \& Bryan 1999) indicate that the injection of turbulence in
clusters is not homogeneous and occurs on very different spatial scales in
different regions; ii) detailed calculations of particle acceleration
due to Alfv\`en waves show that, under reasonable assumptions, the
acceleration efficiency slightly increases with distance from the
cluster center (Brunetti et al. 2004); iii) the radiative losses of
electrons in the innermost cluster regions may be strongly increased
if the magnetic field is larger than the equipartition value, in
particular if it is of the order of B$_{IC}$ (e.g Kuo et al. 2003).

\section{Conclusions}

The link between radio halos and cluster merger processes
is confirmed by the halo  spectral behaviour.
Spectral index maps obtained for the two clusters A665 and A2163 
with an angular resolution of the
order of $\sim$ 1\arcmin, show a clumpy distribution with significant
variations, which are indication of a complex shape of the radiating
electron spectrum. This is supportive of halo models invoking the 
reacceleration of relativistic particles.
Regions of flatter spectra appear to trace the geometry of recent
merger activity as suggested by X-ray maps.  These results prove that
radio emitting electrons are gaining energy from the merger event.
There is no evidence of spectral flattening at the location of the hot
shock detected in A665 (Markevitch \& Vikhlinin 2001).  This favours
the scenario that cluster turbulence might be the major mechanism
responsible for the electron reacceleration.

A plot of the halo total spectral index versus the cluster temperature
indicates that hotter clusters tend to host halos with flatter
spectra. This may be understood in the framework of the reacceleration
models, since hottest clusters, being more massive, may host more
violent mergers.

In the undisturbed cluster regions of A665 and A2163, the spectrum
steepens with the distance from the cluster center. This is
interpreted as the result of the combined effect of a radial decrease
of the cluster magnetic field strength and of the spatial distribution
of the reacceleration efficiency.  The profile of the product between
the cluster magnetic field and the reacceleration efficiency, $B
\chi^2$, derived under simple assumptions is flatter than that
predicted in the case of a constant reacceleration and magnetic field
frozen to the cluster thermal gas.  The ongoing violent mergers may
play a crucial role in determining the conditions of the radiating
particles and of the magnetic fields in clusters.

\acknowledgements{}
L.F. thanks Hyesung Kang and Dongsu Ryu for the invitation to 
such a great conference. 

Basic research in radio astronomy at the Naval Research
Laboratory is supported by the Office of Naval Research.

\end{document}